\documentclass[aps,prb,twocolumn,showpacs,showkeys,preprintnumbers,groupedaddress,amsmath,amssymb,]{revtex4-1}

\usepackage{graphicx,latexsym}
\usepackage{dcolumn}
\usepackage{amsmath,amssymb,epsf,bm}
\usepackage{color}

\begin{document}

\title{Universal conductance fluctuations in indium tin oxide nanowires}

\author{Ping-Yu Yang$^1$}
\author{L. Y. Wang$^2$}
\email{luyao.ep88g@nctu.edu.tw}
\author{Yao-Wen Hsu$^1$}
\author{Juhn-Jong Lin$^{1,2,}$}
\email{jjlin@mail.nctu.edu.tw}

\affiliation{$^1$Institute of Physics, National Chiao Tung University, Hsinchu 30010, Taiwan \\
$^2$Department of Electrophysics, National Chiao Tung University, Hsinchu 30010, Taiwan}

\begin{abstract}

Magnetic field dependent universal conductance fluctuations (UCF's) are observed in weakly disordered indium tin oxide nanowires from 0.26 K up to $\sim 25$ K. The fluctuation magnitudes increase with decreasing temperature, reaching a fraction of $e^2/h$ at $T \lesssim 1$ K. The shape of the UCF patterns is found to be very sensitive to thermal cycling of the sample to room temperatures, which induces irreversible impurity reconfigurations. On the other hand, the UCF magnitudes are insensitive to thermal cycling. Our measured temperature dependence of the root-mean-square UCF magnitudes are compared with the existing theory [C. W. J. Beenakker and H. van Houten, Phys. Rev. B \textbf{37}, 6544 (1988)]. A notable discrepancy is found, which seems to imply that the experimental UCF's are not cut off by the thermal diffusion length $L_T$, as would be expected by the theoretical prediction when $L_T < L_\varphi$, where $L_\varphi$ is the electron dephasing length. The approximate electron dephasing length is inferred from the UCF magnitudes and compared with that extracted from the weak-localization magnetoresistance studies. A reasonable semiquantitative agreement is observed.

\end{abstract}

\pacs{73.23.-b, 73.63.Bd, 72.70.+m, 72.15.Rn}

\maketitle

\section{Introduction}

Universal conductance fluctuations (UCF's) are one of the most meaningful manifestations of the quantum-interference electron transport in mesoscopic and nanoscale systems. \cite{mesoscopic,Akkermans,Washburn86} In weakly disordered miniature metals and at low temperatures, the ``aperiodic" UCF patterns are highly reproducible. Those fluctuation patterns are determined by the specific impurity configuration that is ``frozen" in a given sample at a given cooldown. \cite{Lee-prl85,Altshuler-jetp85,Lee-prb87} Under such conditions, one may sweep a magnetic field \cite{Washburn-rpp92,Thornton-prb87} or gate voltage \cite{Licini-prl85,Kaplan-prl86} sufficiently widely to realize statistically distinct subsystems (independent members) of the specific ensemble which embraces the given sample under study. The sample may then be thermally cycled up to room temperatures to induce possible rearrangement of the impurity configuration. If an impurity/disorder reconfiguration should occur, the UCF patterns would significantly or completely alter after the sample is remeasured at low $T$. In this context, the impurities and defects are essentially {\em static} both in space and with time at liquid-helium temperatures. Apart from the mesoscopic metal and semiconductor structures that are fabricated by the top-down electron-beam lithography, \cite{Washburn86,Umbach-prb84,Skocpol-prl86,Lin-jpcm02} the UCF's have recently been studied in bottom-up artificially synthesized nanowires (NW's). \cite{Hansen-prb05,Schapers-InAs,Schapers-InN,Lien-prb11} Besides, the UCF's have been searched in newly developed materials, such as epitaxial ferromagnets, \cite{Wagner-prl06} carbon nanotubes, \cite{Man-prl05} graphene, \cite{Berezovsky-nano10} and topological insulators. \cite{Checkelsky-prl11}

Tin-doped indium oxide (In$_{2-x}$Sn$_x$O$_{3-\delta}$, or so-called ITO) is a metal oxide which exhibits low electrical resistivities $\rho$ ($\sim$ 10$^2$ $\mu \Omega$ cm) \cite{Li-jap04,Guo-apl11} and, in particular, free-carrier-like electronic conduction properties, \cite{Mryasov-prb01} such as a linear diffusive thermopower in the wide $T$ interval 5--300 K. \cite{Wu-jap10} Single-crystalline ITO NW's possess similar metallic characteristics to those of ITO films. \cite{Chiu-nano09} At not too low temperatures, the $\rho$--$T$ behavior can be described by the standard Boltzmann transport equation. Below a few tens degree of K, the quantum-interference weak-localization (WL) and electron-electron interaction effects \cite{Bergmann-pr84,Altshuler-rev87} (among other possible effects \cite{Chiu-nano09}) cause notable corrections to $\rho$ [see the inset of Fig.~\ref{fig2}(b)]. Furthermore, in {\em sufficiently short} ITO NW's, pronounced UCF's can arise at cryogenic temperatures, owing to the absence of classical self-averaging in the sample conductance $G$. \cite{mesoscopic,Akkermans,Washburn86,  Lee-prl85,Altshuler-jetp85,Lee-prb87} In this work, we report our experimental results for three ITO NW's that reveal marked UCF's in sweeping, perpendicular magnetic fields. Two of our NW's had been intentionally thermally cycled to 300 K and then remeasured at liquid-helium temperatures. The shape of the UCF patterns completely altered due to the impurity/disorder reconfigurations induced by the room-temperature  thermal energy $k_BT$, where $k_B$ is the Boltzmann constant. In sharp contrast, the UCF magnitudes remained basically unchanged. Our measured $T$ dependence of the root-mean-square (rms) UCF magnitudes, however, cannot be described by the existing theory. The reason why seems to be associated with the concept of ``thermal averaging" that is brought forward in the current mesoscopic theory. \cite{Lee-prb87,Beenakker-prb88} Under the condition of $k_BT > \hbar/\tau_\varphi$ (which is pertinent to the present study, where $\hbar$ is the normalized Planck's constant, and $\tau_\varphi$ is the electron dephasing time), the canonical UCF theory predicts that the effect of the thermal averaging would lead to an extra $T^{-1/2}$ temperature dependence of the conductance fluctuation magnitudes, in addition to that to result from the $T$ dependence of $\tau_\varphi$. Unexpectedly, such a strong $T^{-1/2}$ temperature dependence is not seen in this work. Further experimental investigations focusing on the temperature characteristics of the UCF magnitudes would be useful to improve our understanding of this issue. In this regard, self-assembled NW's can provide valuable platforms due to their marked UCF phenomena, as compared with those in conventional lithographic metal structures. Empirically, the UCF effect can often be seen in NW's up to above 10 K, \cite{Hansen-prb05,Schapers-InAs,Schapers-InN,Lien-prb11} while it is seen only below 1 K in top-down lithographic metal structures. \cite{Washburn-rpp92,Wagner-prl06,Mohanty-prl03,Beutler-prl87,Rudolph-prb11}

This paper is organized as follows. Section II contains our experimental method. Section III includes our experimental results and theoretical analysis. Our conclusion is given in Sec. IV. Appendix A contains a discussion of the magnetoresistance in the WL effect in one dimension.

\section{Experimental Method}

ITO nanowires were fabricated by the implantation of Sn ions into In$_2$O$_{3-\delta}$ NW's. The In$_2$O$_{3-\delta}$ NW's were grown by the vapor-solid-liquid (VLS) method, as described previously. \cite{Chiu-nano09} The morphology and the cubic bixbyite structure (the prototype structure being Mn$_2$O$_3$ and the space group: Ia3) of the single-crystalline Sn-doped In$_2$O$_{3-\delta}$ NW's were studied by the scanning electron microscopy (SEM) and the transmission electron microscopy (TEM). The nominal composition of our NW's has previously been determined to be In$_{1.912}$Sn$_{0.088}$O$_{3-\delta}$ (Ref. \onlinecite{Hsu-prb10}).

Four-probe single NW devices were fabricated by the electron-beam lithography. [Figure~\ref{fig2}(a) shows an SEM image of the NW14 nanowire device.] The magnetoresistance (MR) measurements were performed on an Oxford Heliox $^3$He cryostat equipped with a 4-T superconducting magnet. A Linear Research LR-700 ac resistance bridge operating at a frequency of 16 Hz was employed for the MR measurements. To avoid electron heating, an excitation current of $\simeq$ 10 nA (so that the voltage drop along the NW was $\lesssim k_BT/e$, where $e$ is the electron charge) was applied. In all cases, the magnetic field $B$ was applied {\em perpendicular} to the NW axis.

%\begin{table*}
\begin{table}
\caption{Sample parameters for ITO nanowires. $d$ is the diameter, $L$ is the voltage probe distance in a four-probe geometry, $D$ is the electron diffusion constant, $\ell$ is the electron elastic mean free path, and $k_F$ is the Fermi wave number. The uncertainty in $d$ is $\approx \pm5$ nm. $D$, $l$, and $k_Fl$ are for 10 K. The samples are labeled according to their voltage probe distance. The NW28 nanowire was taken from Ref. \onlinecite{Hsu-prb10}.}

\begin{ruledtabular}
\begin{tabular}[t]{lccccccc}

Nanowire & $d$ & $L$ & $\rho$(300 K) & $\rho$(10 K) & $D$ & $\ell$ & $k_Fl$ \\
& (nm) & ($\mu$m) & ($\mu\Omega$ cm) & ($\mu\Omega$ cm) & (cm$^2$/s) & (nm) & \\ \hline

NW12 & 110 & 1.2 & 803 & 740 & 6.6 & 2.8 & 6.8 \\
NW14  & 78 & 1.4  & 576 & 546 & 8.7  & 3.7 & 9.0 \\
NW28  & 72 & 2.8  & 997  & 1030 & 5.5 & 2.8 & 5.7 \\

\end{tabular}
\end{ruledtabular}
\end{table}
%\end{table*}

Table I lists the sample parameters of the three ITO NW's studied in this work. In order to investigate the sensitive effects of the impurity reconfigurations on the UCF patterns (``magneto-fingerprints"), we have thermally cycled the NW12 (NW14) nanowire 3 times (twice) from liquid-helium temperatures up to 300 K. Nevertheless, it should be noted that the impurity/disorder reconfigurations had little effect on the average resistivity value $\langle \rho \rangle$ of a given NW. For example, $\langle \rho (10\, {\rm K}) \rangle$ changed from 546 to 543 $\mu\Omega$ cm for the NW14 nanowire after one thermal cycling. For the NW12 nanowire, $ \langle \rho (10\, {\rm K}) \rangle$ changed from 740 (at first cooldown) to 731 (after first thermal cycling), and then to 721 (after second thermal cycling) $\mu\Omega$ cm. Such small changes in the $\langle \rho \rangle $ values suggest that the number of defects (most likely, point defects,  \cite{Chiu-nano09} which cannot be detected under high-resolution TEM) in our NW's should be sufficiently large, so that our measured $\langle \rho \rangle$ values faithfully reflect statistical average values. In other words, while an impurity reconfiguration would significantly alter the shape of the UCF patterns, it only causes a minor modification to the average resistivity value.

\section{Results and Discussion}

Our NW samples studied in this work possess values of $k_F l > 1$ (see Table I), i.e., they fall in the weakly disordered regime, where $k_F$ is the  Fermi wave number, and $l$ is the electron elastic mean free path. The condition $k_Fl >1$ corresponds to the sample conductance $G > e^2/h$, the quantum conductance. Under such circumstance, the quantum-interference UCF phenomena in short NW's can be expected at low temperatures.

\subsection{UCF's in comparatively long ITO nanowire}

\begin{figure}
\includegraphics[width=10.5cm, height=10cm]{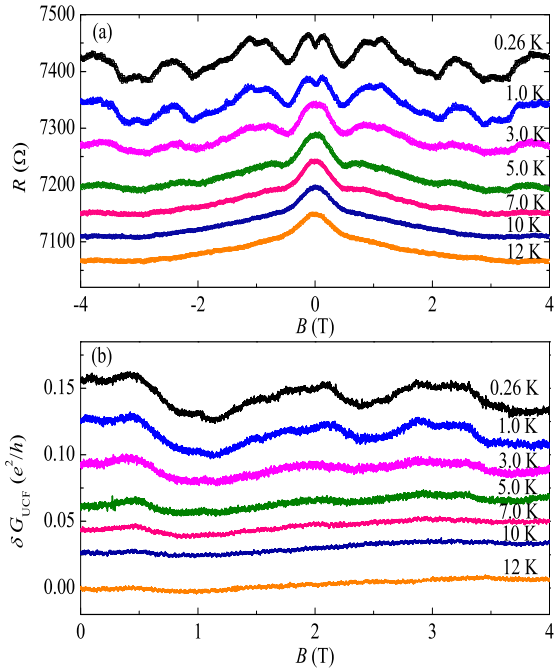}
\caption{(color online) (a) Resistance as a function of magnetic field for NW28 nanowire at several temperatures, as indicated. The MR's are symmetric around $B = 0$. (b) Variation of $\delta G_{\rm UCF}$ with magnetic field for this NW at several $T$ values, as indicated. Note that the $\delta G_{\rm UCF}$ magnitudes increase with reducing $T$. In panels (a) and (b), the curves are vertically offset for clarity. \label{fig1}}
\end{figure}

Figure~\ref{fig1}(a) shows our measured MR's for the NW28 nanowire at several $T$ values, as indicated. Resistance fluctuations are clearly evident, especially at the lowest measurement temperatures. Moreover, it can be seen that the MR's are essentially symmetric about $B = 0$, implying a high contact transparency of our nanowire/lithographic-electrode interface. \cite{Hansen-prb05} These aperiodic, strongly $T$ dependent resistance fluctuations arise from the UCF mechanism, which is the central theme of this paper. In low magnetic fields, the WL/weak-antilocalization contribution to the MR is also present (roughly speaking, the MR's in $|B| < 0.2$ T), which can be quantitatively analyzed according to the standard one-dimensional (1D) theoretical predictions \cite{Altshuler-rev87,Hsu-prb10} (see Appendix A for the formula of the 1D WL MR and our least-squares fits). Therefore, we may write the total conductance at a given $T$ as follows
\begin{equation} \label{MG}
G(B) = G_0 + \delta G_{\rm WL}(B) + \delta G_{\rm UCF}(B) \,,
\end{equation}
where $G_0 = G(B = 0)$, $\delta G_{\rm WL}(B)$ is the magnetoconductance in the 1D WL effect, and $\delta G_{\rm UCF}(B)$ is the 1D UCF contribution. The UCF signals are thus obtained by subtracting the measured $G_0$ and the least-squares fitted $\delta G_{\rm WL}$ from the total $G(B)$, i.e., $\delta G_{\rm UCF}(B) = G(B) - G_0 - \delta G_{\rm WL}(B)$.

Figure~\ref{fig1}(b) shows the variation of $\delta G_{\rm UCF}$ with perpendicular magnetic field at several temperatures, as indicated. This figure illustrates that the UCF magnitudes are progressively suppressed as $T$ increases, and they disappear around $\sim 12$ K. Here  $\delta G_{\rm UCF}$ is plotted in units of $e^2/h$. In this NW, the peak-to-peak value is $\delta G_{\rm UCF}(0.26\, {\rm K}) \sim 0.03 e^2/h$. This value is notably smaller than that ($\sim e^2/h$) as would be expected for a 1D mesoscopic sample at absolute zero. \cite{Lee-prl85,Altshuler-jetp85,Lee-prb87} This is partly because this NW has a sample length of $L \approx 2.8$ $\mu$m ($L$ is the voltage probe distance in a four-probe geometry), which is almost 20 times its electron dephasing length $L_\varphi (0.26\, {\rm K}) \approx 170$ nm. \cite{Hsu-prb10} Therefore, the effect of classical self-averaging over {\em independent} phase-coherence segments has greatly suppressed the measured $\delta G_{\rm UCF}$ magnitudes of the entire sample. In order to augment the entire NW UCF magnitudes to allow more quantitative analysis, we have focused our measurements particularly on the two NW12 and NW14 nanowires, which are intentionally made to retain small $L/L_\varphi (0.26 \, {\rm K})$ ratio values so that the ensemble-averaging effect is largely minimized.

We would like to  note in passing that the UCF signals in conventional lithographic mesoscopic metal structures are generally observed only at $T < 1$ K. \cite{Washburn-rpp92,Wagner-prl06,Mohanty-prl03,Beutler-prl87,Rudolph-prb11} On the contrary, the UCF's are found to persist up to (far) above 10 K in a number of bottom-up artificially synthesized NW's. \cite{Hansen-prb05,Schapers-InAs,Schapers-InN,Lien-prb11} This implies that many as-grown single-crystalline (both metallic  \cite{Chiu-nano09} and semiconducting \cite{Chiu-nano09ZnO}) NW's must contain high levels of point defects, which facilitated pronounced diffusive electron motion.

\subsection{UCF's in short ITO nanowires}

\begin{figure}
\includegraphics[width=10.5cm, height=10cm]{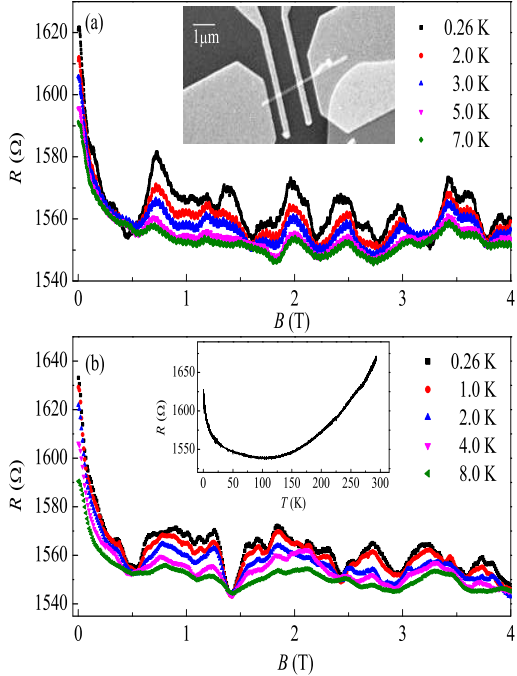}
\caption{(color online) Resistance fluctuations at several $T$ values in NW14 nanowire (a) at first cooldown (from top down: 0.26, 2.0, 3.0, 5.0, and 7.0 K), and (b) after one thermal cycling to 300 K (from top down: 0.26, 1.0, 2.0, 4.0, and 8.0 K). The inset in panel (a) shows an SEM image of this single NW device. The inset in panel (b) plots the resistance versus temperature for this cooldown. Note that the resistance values maintain similar in panels (a) and (b), while the fluctuation patterns (``magneto-fingerprints") become uncorrelated. \label{fig2}}
\end{figure}

Figure~\ref{fig2}(a) and \ref{fig2}(b) show the raw resistance as a function of magnetic field for the NW14 nanowire at first cooldown and after one thermal cycling to room temperatures, respectively. That is, to perform the UCF measurements, we first cooled the NW down to 0.26 K and subsequently measured the MR curves at several selected temperatures between 0.26 and 10 K. The MR curves were measured in sequence with progressive increases in $T$. This set of MR curves is shown in Fig.~\ref{fig2}(a). Then, we warmed the NW up to 300 K, staying overnight to allow possible thermal energy induced impurity reconfiguration, and cooled down the NW again to 0.26 K. A second series of MR curves were then measured with gradual increases in $T$.  This set of MR curves is plotted in Fig.~\ref{fig2}(b). [The inset of Fig.~\ref{fig2}(b) plots the corresponding resistance versus temperature data for this cooldown.]

These two figures clearly reveal the WL-induced MR's in low magnetic fields (the sharp resistance drops in $B \lesssim 0.1$ T, see Appendix A) as well as the UCF signals in higher magnetic fields. In particular, it can be seen that the shape of the UCF patterns remains essentially unchanged in a given run, albeit the magnitudes decrease with increasing $T$. What is even more significant is that the UCF patterns completely altered after the sample was thermally cycled to 300 K and then remeasured at low temperatures. Nevertheless, the corresponding UCF magnitudes remained similar before and after the thermal cycling. The profound change in the shape of the UCF patterns can be readily ascribed to an impurity/disorder reconfiguration as a consequence of the warmup to 300 K. On the other hand, the similar UCF magnitudes between the two runs can be understood in terms of a similar electron dephasing length $L_\varphi$ ($L_\varphi^{\rm UCF}$) at a given $T$. That is, the size of $L_\varphi$ ($L_\varphi^{\rm UCF}$) is essentially determined by the amount of impurity or the degree of disorder in the NW, which was essentially unaffected by thermal cycling. (For the convenience of the following discussion, we shall use $L_\varphi$ to denote the electron dephasing length extracted from the WL MR studies, while using $L_\varphi^{\rm UCF}$ to denote that inferred from the UCF measurements.)

\begin{figure}
\includegraphics[width=10cm, height=10cm]{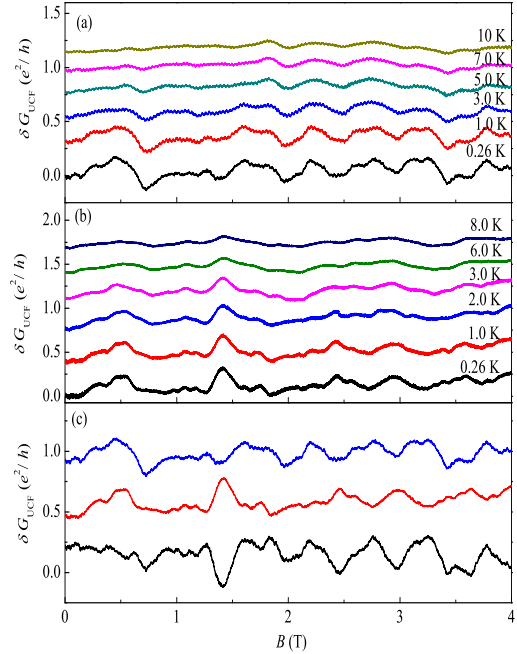}
\caption{(color online) Variation of $\delta G_{\rm UCF}$ with magnetic field at several $T$ values for NW14 nanowire (a) at first cooldown, and (b) after one thermal cycling to 300 K. (c) The $\delta G_{\rm UCF}$(0.26\,K) curves taken from panel (a) (top curve) and panel (b) (middle curve), and their difference (bottom curve), as a function of magnetic field. In panels (a) to (c), the curves are vertically offset for clarity. \label{fig3}}
\end{figure}

According to Eq.~(\ref{MG}), we have calculated $\delta G_{\rm UCF}$ [by subtracting the measured $G_0$ and the least-squares fitted $\delta G_{\rm WL}(B)$ from the total $G(B) = 1/R(B)$] for the NW14 nanowire and plotted in Figs.~\ref{fig3}(a) and \ref{fig3}(b) the variation of $\delta G_{\rm UCF}$ with $B$ at first cooldown and after one thermal cycling to room temperatures, respectively. Inspection of these two panels indicates that, in both runs, the peak-to-peak $\delta G_{\rm UCF}$(0.26\,K) magnitudes are  $\approx 0.3 e^2/h$. This magnitude is close to the theoretically expected amplitude of $\sim e^2/h$.

\begin{figure}
\includegraphics[width=10cm, height=10cm]{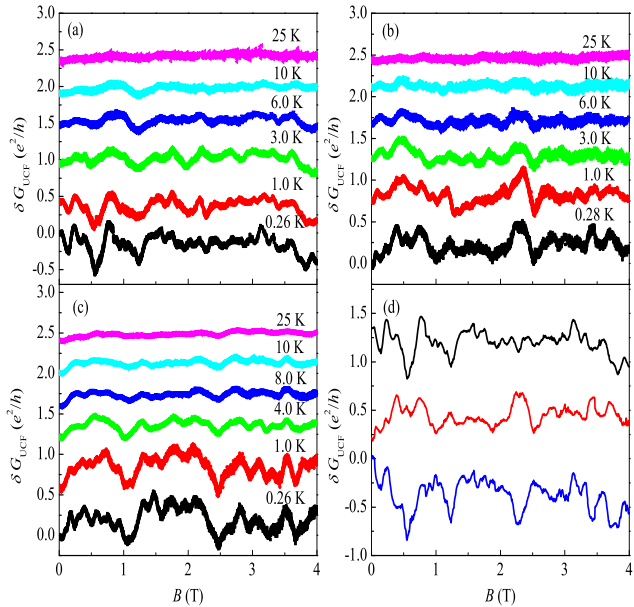}
\caption{(color online) Variation of $\delta G_{\rm UCF}$ with magnetic field at several $T$ values for NW12 nanowire (a) at first cooldown, (b) after first thermal cycling, and (c) after second thermal cycling to 300 K. (d) The $\delta G_{\rm UCF}$(0.26\,K) curves taken from panel (a) (top curve) and panel (b) (middle curve), and their difference (bottom curve), as a function of magnetic field. In panels (a) to (d), the curves are vertically offset for clarity. \label{fig4}}
\end{figure}

Figure~\ref{fig3}(c) shows a plot of the two measured $\delta G_{\rm UCF}(B)$ curves at 0.26 K for the NW14 nanowire at first cooldown (top curve) and after one thermal cycling to room temperatures (middle curve), together with their difference (bottom curve). This figure clearly manifests that the UCF patterns are {\em uncorrelated} between the two runs, and that their difference possesses a magnitude similar to the magnitude ($\approx 0.3 e^2/h$) in each run. This observation unambiguously demonstrates the nature of the sensitivity of the UCF patterns to the specific impurity configuration.

The genuine UCF characteristics in short ITO NW's are further confirmed by examining one additional NW. Figures~\ref{fig4}(a), \ref{fig4}(b) and \ref{fig4}(c) show the variation of $\delta G_{\rm UCF}$ with $B$ at several temperatures for the NW12 nanowire at first cooldown, after first thermal cycling, and after second thermal cycling to room temperatures, respectively. The major UCF features, such as the increased fluctuation magnitudes with decreasing $T$, are clearly seen. In particular, the peak-to-peak $\delta G_{\rm UCF}$(0.26 K) magnitudes reach $\approx 0.5 e^2/h$. As $T$ increases, the UCF signals become totally suppressed around 25 K, a relatively high $T$ value compared with that ($< 1$ K) usually seen in conventional lithographic mesoscopic metal samples. Figure~\ref{fig4}(d) shows the $\delta G_{\rm UCF}$(0.26\,K) curves taken from Figs.~\ref{fig4}(a) (top curve) and \ref{fig4}(b) (middle curve). The bottom curve is a plot of the difference between these two curves. It is evident that the difference also fluctuates with peak-to-peak magnitudes of $\approx 0.5 e^2/h$, strongly demonstrating that these two $\delta G_{\rm UCF}$(0.26\,K) curves are uncorrelated.

In brief, our observations in Figs.~\ref{fig2} to \ref{fig4} illustrate that the shape of the UCF patterns is very sensitive to the specific impurity configuration in a given NW at a given cooldown. On the other hand, the UCF magnitudes in a given NW are only sensitive to the measurement temperature, which determines the size of $L_\varphi^{\rm UCF}$ ($L_\varphi$).

\subsection{Comparison with theory and the problem with thermal averaging}

\begin{figure}
\includegraphics[width=9cm, height=8cm]{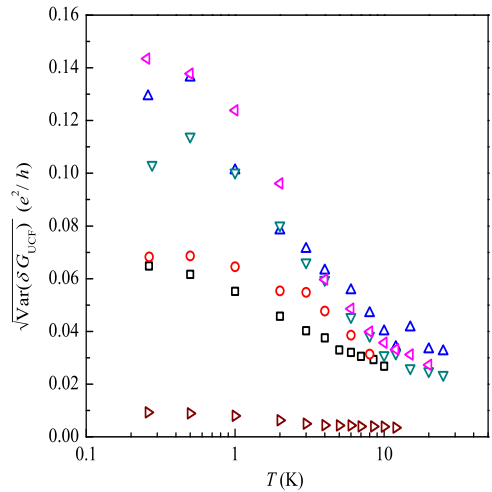}
\caption{(color online) Variation of $\sqrt{{\rm Var}(\delta G_{\rm UCF})}$ with temperature for NW12, NW14, and NW28 nanowires. The NW12 nanowire had been measured 3 times: at first cooldown ($\triangle$), after first thermal cycling ($\triangledown$), and after second thermal cycling to 300 K ($\triangleleft$). The NW14 nanowire had been measured twice: at first cooldown ($\Box$), and after one thermal cycling to 300 K ($\circ$). The NW28 nanowire had been measured at first cooldown ($\triangleright$). Note that the $\sqrt{{\rm Var}(\delta G_{\rm UCF})}$ magnitudes vary greatly from sample to sample, but they are not sensitive to thermal cycling to 300 K for a given NW. \label{fig5}}
\end{figure}

To quantitatively compare the measured UCF magnitudes with theoretical predictions, we plot $\sqrt{{\rm Var}(\delta G_{\rm UCF})}$ as a function of temperature for our three ITO NW's in Fig.~\ref{fig5}. Here the variance of the UCF magnitudes is defined by \cite{Lee-prb87}
\begin{equation}
{\rm Var}(\delta G_{\rm UCF}) =  \big{\langle} \bigl[ \delta G_{\rm UCF} (B) - \langle \delta G_{\rm UCF} (B) \rangle \bigr] ^2 \big{\rangle} \,, \label{variance}
\end{equation}
where $\langle ... \rangle$ denotes an average over the magnetic field. This is equivalent to an ensemble average over impurity configurations, according to the ``ergodic hypothesis" assumed by Lee, Stone and Fukuyama. \cite{Lee-prb87} Note that the NW12 (NW14) nanowire has been measured 3 times (twice). Figure~\ref{fig5} indicates that the $\sqrt{{\rm Var}(\delta G_{\rm UCF})}$ magnitude decreases with increasing $T$ in every measurement. At 0.26 K, $\sqrt{{\rm Var}(\delta G_{\rm UCF})} \approx 0.14 e^2/h$, $\approx 0.07 e^2/h$, and $\approx 0.01 e^2/h$ for the NW12, NW14, and NW28 nanowires, respectively. The large differences among these magnitudes suggest that the UCF phenomena are highly sensitive to specific samples (and that the fluctuations are not due to instrumental noises). It should be stressed again that our measured $\sqrt{{\rm Var}(\delta G_{\rm UCF})}$ magnitudes maintain essentially similar for a given NW, irrespective to thermal cycling. In order to achieve a good understanding of the microscopic UCF physics, it is desirable to explain not only the low-temperature value but also the $T$ dependence of the $\sqrt{{\rm Var}(\delta G_{\rm UCF})}$ magnitude, as we carry out below.

The UCF theory predicts a fluctuation magnitude of $0.73 e^2/h$ for a weakly disordered, 1D mesoscopic wire ($l \ll L \lesssim L_\varphi$) at $T = 0$ K and in zero magnetic field. \cite{Lee-prb87,Beenakker-prb88} In the presence of a sufficiently large magnetic field $|B| > B_c$, where $B_c$ is the correlation field (see below), the UCF magnitude should be suppressed by a factor of $1/\sqrt{2}$ (Refs. \onlinecite{Lee-prb87,Beenakker-prb88,Stone-prb89}). This is because a magnetic field breaks the time-reversal symmetry of the Cooperon (particle-particle) propagator, leaving the diffuson (particle-hole) propagator as the remaining contribution to the UCF effect. Therefore, the saturated rms fluctuation magnitude in 1D would be $\approx \frac{1}{\sqrt{2}} \times 0.73 e^2/h \approx 0.5 e^2/h$. Taking the NW12 nanowire as an example, we have obtained $L_\varphi (0.26 \, {\rm K}) \approx 350$ nm from the WL MR studies [Fig.~\ref{fig8}(b)]. Since this NW has a sample length $L \simeq 1.2$ $\mu$m, a quick estimate gives a theoretical magnitude of $\sim \frac{1}{\sqrt{N}} \times 0.5 e^2/h \simeq 0.27 e^2/h$, where $N \simeq L/L_\varphi (0.26\, {\rm K})$ is the number of independent phase-coherence regions. Thus, the experimental result agrees satisfactorily with the theoretical prediction to within a factor of $\sim 2$ in this limit.

At $T > 0$ K, in addition to the classical self-averaging effect due to reduced $L_\varphi^{\rm UCF}$ with increasing $T$, the thermal averaging effect would need to be taken into consideration when $k_BT > \hbar/\tau_\varphi$, where $\tau_\varphi (T)$ is the electron dephasing time. \cite{Lee-prb87} Under this condition, the number of uncorrelated energy regimes involved in the quantum-interference electron transport is $N_c \simeq (k_BT)/(\hbar /\tau_\varphi) = (L_\varphi^{\rm UCF} /L_T)^2$, where $L_T = \sqrt{D\hbar /k_BT}$ is the thermal diffusion length, and $D$ is the electron diffusion constant. Consequently, a quantitative description of the temperature dependence of the $\sqrt{{\rm Var}(\delta G_{\rm UCF})}$ magnitude becomes a challenging task. This has to be solved by explicitly calculating the conductance autocorrelation function, which is defined by \cite{Lee-prl85,Lee-prb87, Altshuler-jetp85a,Beenakker-prb88}
\begin{widetext}
\begin{equation}
F(\triangle B) = \big{\langle} \big[ \delta G_{\rm UCF}(B) - \langle \delta G_{\rm UCF} (B) \rangle \big] \bigl[ \delta G_{\rm UCF}(B+\triangle B) - \langle \delta G_{\rm UCF}(B + \triangle B) \rangle \big] \big{\rangle} \,, \label{F(B)}
\end{equation}
\end{widetext}
where $\langle ... \rangle$ denotes an average over the magnetic field. Again, this is equivalent to an ensemble average over impurity configurations. It should be noted that $F(\triangle B)$ depends only on the difference in magnetic field $\triangle B$, but not on $B$ itself for $|B| \gtrsim B_c$. Therefore, unlike the WL effect, the UCF phenomena (from the diffuson channel) can persist up to relatively high magnetic fields (e.g., $\sim$ 10 T) in mesoscopic metal structures. \cite{Washburn86} By comparing Eqs.~(\ref{variance}) and (\ref{F(B)}), one immediately sees that $F(0) = {\rm Var}(\delta G_{\rm UCF})$.

In the microscopic theory of Lee, Stone and Fukuyama, \cite{Lee-prb87} Eq.~(\ref{F(B)}) is expressed in terms of an integral, which can be evaluated analytically only in the asymptotic regimes of $L_\varphi^{\rm UCF} \ll L_T$ and $L_T \ll L_\varphi^{\rm UCF}$. In experiments, however, these two characteristic length scales are often comparable, namely, $L_\varphi^{\rm UCF} \sim L_T$. In order to facilitate comparison with the 1D experiment ($l \ll d < L_T, L_\varphi^{\rm UCF} < L$, where $d$ is the diameter of the NW), Beenakker and van Houten have proposed an approximate formula (accurate to within 10\%) to interpolate between the two asymptotic regimes: \cite{Beenakker-prb88}
\begin{equation}
F(0) \simeq \alpha \biggl( \frac{e^2}{h} \biggr)^2 \biggl( \frac{L_\varphi^{\rm UCF}}{L} \biggr)^3 \biggl[ 1 + \frac{\alpha}{\beta} \biggl( \frac{L_\varphi^{\rm UCF}}{L_T} \biggr)^2 \biggr]^{-1} \,, \label{F(0)}
\end{equation}
where the numerical prefactors in the presence of $|B| > B_c$ are $\alpha = 6$ and $\beta = \frac43 \pi$.  Note that Eq.~(\ref{F(0)}) recovers the asymptotic results of $F(0) \simeq \alpha (e^2/h)^2 (L_\varphi^{\rm UCF}/L)^3$ (for $L_\varphi^{\rm UCF} \ll L_T$) and $F(0) \simeq \beta (e^2/h)^2 (L_T^2 L_\varphi^{\rm UCF} /L^3)$ (for $L_T \ll L_\varphi^{\rm UCF}$), which were originally obtained in Ref. \onlinecite{Lee-prb87}. This formula has recently been applied to explain the $T$ dependence of the rms UCF magnitudes in InAs (Ref. \onlinecite{Schapers-InAs}) and InN (Ref. \onlinecite{Schapers-InN}) NW's, but the authors of Refs. \onlinecite{Schapers-InAs} and \onlinecite{Schapers-InN} had to treat $\alpha$ as a fitting parameter in order to bring the theoretical values to be close to the experimental values. In the present work, surprisingly, we find that Eq.~(\ref{F(0)}) can not even explain our experimental results in a qualitative manner.

Experimentally, we have observed that  the $\sqrt{{\rm Var}(\delta G_{\rm UCF})}$ magnitudes are nearly temperature independent below $\sim 2$ K. That is, there is a tendency toward a saturation of the rms UCF magnitudes at low temperatures (Fig.~\ref{fig5}). On the contrary, according to Eq.~(\ref{F(0)}) and in the limit of $L_T < L_\varphi^{\rm UCF} < L$ (which applies to our experimental situation), one should expect an approximate $\sqrt{F(0)} = \sqrt{{\rm Var}(\delta G_{\rm UCF})} \propto (L_\varphi^{\rm UCF})^{1/2} L_T \propto T^{-1/2}$ temperature dependence. Here the $T$ dependence of $({L_\varphi^{\rm UCF}})^{1/2}$ due to, e.g., the 1D Nyquist quasielastic electron-electron ($e$-$e$) scattering time ($\tau_{ee}^N \propto T^{-2/3}$), \cite{Altshuler-jpc82} is comparatively weak and may be ignored for the purpose of our discussion. Therefore, without performing any quantitative comparison, we can already rule out the possibility of applying Eq.~(\ref{F(0)}) to describe the $T$ dependence of our measured $\sqrt{{\rm Var}(\delta G_{\rm UCF})}$ magnitudes.

\begin{figure}
\includegraphics[width=10cm, height=8cm]{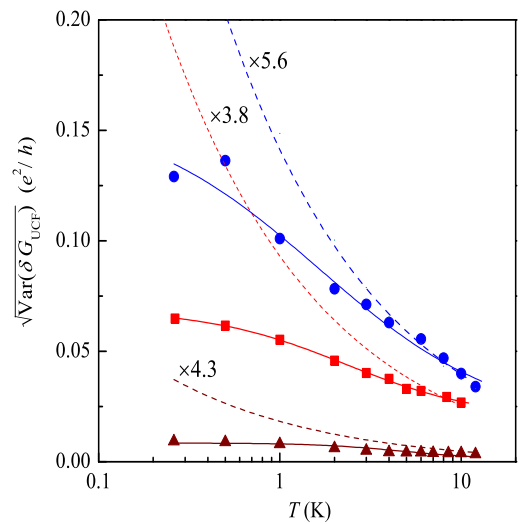}
\caption{(color online) Variation of measured $\sqrt{{\rm Var}(\delta G_{\rm UCF})}$ magnitudes with temperature for the NW12 (circles), NW14 (squares), and NW28 (triangles) nanowires at first cooldown. The solid curves drawn through the data points are guides to the eye. The dashed curves are the theoretical predictions of Eq.~(\ref{F(0)}) evaluated by substituting the corresponding measured $L_\varphi$ values. The theoretical and experimental values are normalized for 10 K in each NW, by multiplying the theoretical values by factors of 5.6, 3.8, and 4.3 for NW12, NW14, and NW28 nanowires, respectively. \label{fig6}}
\end{figure}

Figure~\ref{fig6} plots our measured $\sqrt{{\rm Var}(\delta G_{\rm UCF})}$ magnitudes as a function of temperature for the NW12 (circles), NW14 (squares), and NW28 (triangles) nanowires at first cooldown. Also plotted are the theoretical predictions of Eq.~(\ref{F(0)}) (dashed curves). Note that the theoretical predictions are plotted by substituting the corresponding $L_\varphi$ values extracted from the WL MR studies. (The estimate of $L_\varphi^{\rm UCF}$ values from the UCF effect is to be discussed below.) For the convenience of comparison, the theoretical and experimental values are normalized for 10 K. This has been done by multiplying the theoretical values by a factor of $\sim 5$ in all three NW's, as indicated in the main panel of Fig.~\ref{fig6}. This figure shows the divergences between the experiment and theory at low $T$. Our observation of nearly saturated $\sqrt{{\rm Var}(\delta G_{\rm UCF})}$ magnitudes at $T \lesssim 2$ K suggests that the phase-coherence region  in our NW's is not cut off by $L_T$ (see also Ref. \onlinecite{thermal}). The reason why is not understood at present and should deserve further investigations.

\subsection{Electron dephasing length}

While Eq.~(\ref{F(0)}) does not describe the $T$ dependence of the $\sqrt{{\rm Var}(\delta G_{\rm UCF})}$ magnitude satisfactorily, we may still apply Eq.~(\ref{F(B)}) to estimate the semiquantitative $L_\varphi^{\rm UCF}$ values from the UCF signals. By definition, the correlation field $B_c$ is the characteristic magnetic field corresponding to half maximum of the autocorrelation function $F(\triangle B = B_c)  =  \frac12  F(0)$. Heuristically, $B_c$ defines a (perpendicular) field scale such that the magnetic flux enclosed by a phase-coherence segment of the NW satisfies the relation \cite{Lee-prb87,Beenakker-prb88}
\begin{equation}
B_c L_\varphi^{\rm UCF} d \simeq \tilde\gamma (T)\, \frac{h}{e} \,, \label{gamma}
\end{equation}
where $h/e$ is the flux quantum, and $\tilde\gamma$ is a numerical prefactor. In other words, $B_c$ represents the typical scale of the spacing between peaks and valleys in the conductance fluctuations. Thus, $L_{\varphi}^{\rm UCF} (T)$ may (and, according to the existing theory, can only) be calculated through the measured $B_c (T)$, if $\tilde\gamma (T)$ is known.

\begin{figure}
\includegraphics[width=9cm, height=8cm]{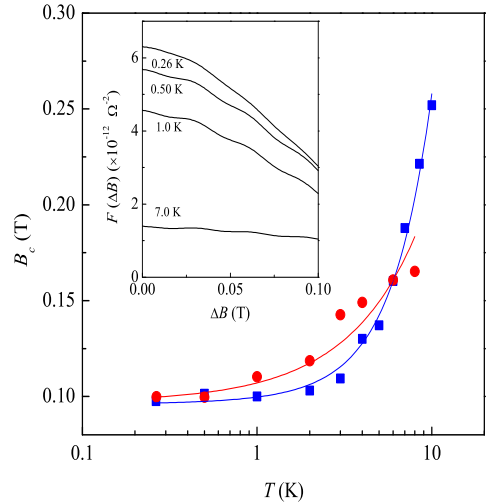}
\caption{(color online) Variation of correlation magnetic field with temperature for NW14 nanowire at first cooldown (squares), and after one thermal cycling to 300 K (circles). The solid curves drawn through the data points are guides to the eye. Inset: Conductance autocorrelation function at four $T$ values, as indicated, for the same NW at first cooldown. \label{fig7}}
\end{figure}

\begin{figure}
\includegraphics[width=10cm, height=10cm]{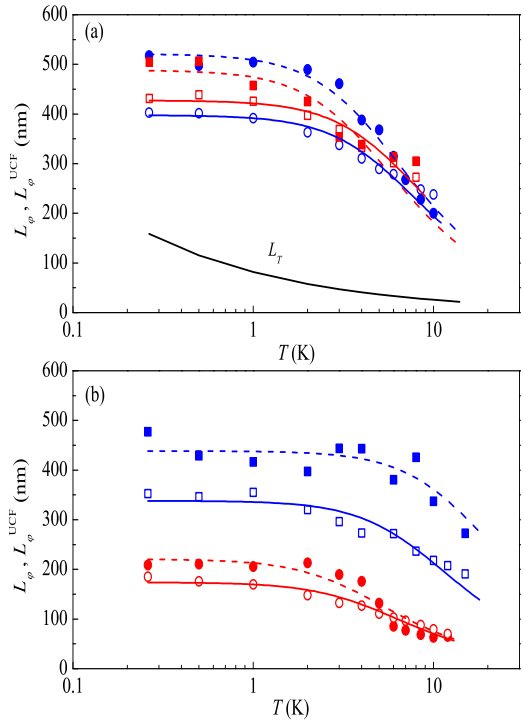}
\caption{(color online) (a) Variation of $L_\varphi^{\rm UCF}$ and $L_\varphi$ with temperature for NW14 nanowire: $L_\varphi^{\rm UCF}$ at first cooldown (closed circles), and after one thermal cycling to 300 K (closed squares); and $L_\varphi$ at first cooldown (open circles), and after one thermal cycling to 300 K (open squares). The bottom curve shows $L_T$ calculated for the first cooldown. (b) Variation of $L_\varphi^{\rm UCF}$ and $L_\varphi$ with temperature for NW12 and NW28 nanowires at first cooldown: $L_\varphi^{\rm UCF}$ (closed squares) and $L_\varphi$ (open squares) for NW12 nanowire; and $L_\varphi^{\rm UCF}$ (closed circles) and $L_\varphi$ (open circles) for NW28 nanowire. In panels (a) and (b), the solid curves drawn through $L_\varphi$ are least-squares fits to Eq.~(\ref{tauphi}), while the dashed curves drawn through $L_\varphi^{\rm UCF}$ are guides to the eye. \label{fig8}}
\end{figure}

The value of $\tilde\gamma$ depends on the relative size of $L_\varphi^{\rm UCF}$ to $L_T$. Since $L_\varphi^{\rm UCF}$ and $L_T$ generally possess different temperature dependencies, $\tilde\gamma$ is a complex function of $T$. Its value has been calculated analytically only for the asymptotic regimes: $\tilde\gamma \simeq 0.95$ (for $L_T \ll L_\varphi^{\rm UCF}$) and $\simeq 0.42$ (for $L_\varphi^{\rm UCF} \ll L_T$). \cite{Lee-prb87,Beenakker-prb88} In experiments, as mentioned, these two characteristic length scales are often comparable. In this work, we find that $L_\varphi^{\rm UCF}$ is a few times longer than $L_T$ in our NW's [see Fig.~\ref{fig8}(a)]. Therefore, we may tentatively substitute $\tilde\gamma \simeq 0.95$ into Eq.~(\ref{gamma}) to compute the approximate values of $L_\varphi^{\rm UCF}$ using our measured $B_c$.

Figure~\ref{fig7} plots our extracted $B_c$ as a function of temperature for the NW14 nanowire at first cooldown (squares) and after one thermal cycling to room temperatures (circles). The size of a $B_c \sim 0.1$--0.2 T suggests that our experimental rms UCF magnitudes (Fig.~\ref{fig5}) were deduced from averaging over $\sim 20$ to 40 $B_c$ periods for a measuring magnetic field of 4 T. The inset shows the corresponding $F(\triangle B)$ at several $T$ values for the same NW at first cooldown. It should be stressed that the values of $B_c$, which are empirically extracted from $F(\triangle B = B_c) = \frac12 F(0)$, depend only on the definition of the conductance autocorrelation function $F(\triangle B)$, Eq.~(\ref{F(B)}), but  not on the specific functional form of the Eq.~(\ref{F(0)}). Therefore, it is justified to use the Eq.~(\ref{gamma}) to estimate the $L_\varphi^{\rm UCF}$ values. The errors in such estimates would then arise mainly from the uncertainties in the numerical value of $\tilde\gamma$ (and in the NW diameter $d$). Nevertheless, $\tilde\gamma$ should be of order unity, because the physical meaning of the Eq.~(\ref{gamma}) is transparent, namely, $B_c$ corresponds to the field scale that leads to a threading magnetic flux (approximately) equal to one flux quantum $h/e$ in a phase-coherence region of the NW.

Figure~\ref{fig8}(a) shows a plot of the extracted electron dephasing length $L_\varphi^{\rm UCF}$, along with the $L_\varphi$ inferred from the WL MR studies (see Appendix A), as a function of $T$ for the NW14 nanowire both at first cooldown and after one thermal cycling to room temperatures. The corresponding $L_T$ calculated from the first cooldown is also plotted for comparison. This figure indicates that $L_\varphi^{\rm UCF}$ lies slightly above $L_\varphi$, as it should be, since our evaluated $L_\varphi^{\rm UCF}$ represents the upper bound of the dephasing length defined in Eq.~(\ref{gamma}). For instance, $L_\phi^{\rm UCF}$(0.26 K) is about 20\% higher than $L_\varphi$(0.26 K). Such a level of agreement is satisfactory. \cite{uncertainty} Similarly, Fig.~\ref{fig8}(b) plots the variation of $L_\varphi^{\rm UCF}$ and $L_\varphi$ with temperature for the two NW12 and NW28 nanowires at first cooldown, as indicated in the caption to Fig.~\ref{fig8}. It can be seen that $L_\varphi^{\rm UCF}$ lies above, but close to, its corresponding $L_\varphi$ in each NW. This observation suggests that substituting $\tilde\gamma \simeq 0.95$ in  Eq.~(\ref{gamma}) can provide a reasonable, although not exact, estimate for $L_\varphi^{\rm UCF}$ in our NW's (Ref. \onlinecite{scatter}).

In the ITO material, the microscopic electron dephasing processes have recently been identified, and the total dephasing rate is found to be given by \cite{Wu-prb}
\begin{equation}
\frac{1}{\tau_\varphi (T)} = \frac{1}{\tau_0} + \frac{1}{\tau_{ee}^N (T)} + \frac{1}{\tau_{ee}(T)} \,, \label{tauphi}
\end{equation}
where $1/\tau_0$ is a constant (or a very weakly $T$ dependent dephasing process), \cite{Lin-jpcm02,Lin-prb87} $1/\tau_{ee}^N = A_{ee}^N T^{2/3}$ is the the 1D small-energy-transfer $e$-$e$ scattering rate, \cite{Altshuler-jpc82} and $1/\tau_{ee} = A_{ee} T^2 \, {\rm ln}(E_F/k_BT)$ is the large-energy-transfer $e$-$e$ scattering rate, \cite{Fukuyama-prb83} where $E_F$ is the Fermi energy. The quasielastic Nyquist term $1/\tau_{ee}^N$ should dominate in a wide $T$ interval at liquid-helium temperatures, while the $1/\tau_{ee}$ term would dominate only until several tens degree of K in this particular material. (For reference, $A_{ee} \sim 1 \times 10^7$ K$^{-2}$ s$^{-1}$ in ITO, see Ref. \onlinecite{Wu-prb}.) Therefore, the $1/\tau_{ee}$ term may be ignored if we focus on $T \lesssim 20$ K. \cite{e-ph}

\begin{table}
\caption{Adjustable parameters for $1/\tau_\varphi$, Eq.~(\ref{tauphi}), at first cooldown. $1/\tau_0$ is in s$^{-1}$, and $A_{ee}^N$ and $(A_{ee}^N)^{th}$ are in K$^{-2/3}$ s$^{-1}$.}

\begin{ruledtabular}
\begin{tabular}[t]{lccccc}

Nanowire & $1/\tau_0$ & $A_{ee}^N$ & $(A_{ee}^N)^{th}$ \\ \hline
NW12  & 4.3$\times$10$^9$ & 1.4$\times$10$^9$  & 3.3$\times$10$^9$ \\
NW14  & 4.3$\times$10$^9$ & 1.5$\times$10$^9$  & 5.3$\times$10$^9$ \\

\end{tabular}
\end{ruledtabular}
\end{table}

For 1D quasielastic Nyquist $e$-$e$ scattering, the theory \cite{Hsu-prb10,Altshuler-jpc82} predicts a coupling strength $(A_{ee}^N)^{th} = [(e^2 \sqrt{D}R k_B) / (2 \sqrt{2} \hbar^2 L)]^{2/3}$, where $R$ is the resistance of the NW, and $L$ is the NW segment between the two voltage probes. We have compared our experimental $L_\varphi$ (but not $L_\varphi^{\rm UCF}$) with Eq.~(\ref{tauphi}), and our least-squares fitted values of the adjustable parameters $1/\tau_0$ and $A_{ee}^N$ are listed in Table II. It can be seen that our experimental value of $A_{ee}^N$ agrees with the theoretical value $(A_{ee}^N)^{th}$ to within a factor of $\sim 2$ ($\sim 3$) for the NW12 (NW14) nanowire. These results provide meaningful self-consistency check of our experimental method and data analyses.

\section{Conclusion}

We have observed universal conductance fluctuations with varying magnetic field in indium tin oxide nanowires from 0.26 K up to $\sim 25$ K. The UCF's originate from the inherent quantum-interference nature of the electron transport in weakly disordered nanoscale structures. We found that the shape of the UCF patterns is very sensitive to the specific impurity configuration, and it alters completely after thermal cycling the sample to room temperatures. The root-mean-square UCF magnitudes increase with reducing temperature, reaching a fraction of $e^2/h$ at $T \lesssim 1$ K. However, the temperature dependence of our measured UCF magnitudes cannot be explained by the existing theory. The discrepancy between the experiment and theory seems to arise from the absence of the thermal averaging effect in our measurements. In our experiment, we are always in the regime of $k_BT > \hbar/\tau_\varphi$, corresponding to $L_T < L_\varphi$ ($L_\varphi^{\rm UCF}$). Under such conditions, one would expect the responsible phase-coherence region in our nanowires to be cut off by $L_T$, instead of by $L_\varphi$ ($L_\varphi^{\rm UCF}$). The reason why this does not happen such that $L_T \propto T^{-1/2}$ does not lead to a notable temperature dependence of the measured UCF magnitudes is not understood. In this work, the UCF magnitudes in every nanowire are deduced from averaging over about 20 to 40 correlation field $B_c$ periods for our applied magnetic field of 4 T. Whether such averaging over a somewhat limited range of measuring magnetic field is fully equivalent to the theoretically concerned averaging over a complete change of the impurity configurations, i.e., whether the ergodic hypothesis assumed in the original UCF theory \cite{Lee-prl85,Lee-prb87} is faithfully met in our measurements, deserves further investigations. Finally, the approximate values of the electron dephasing length $L_\varphi^{\rm UCF}$ has been evaluated and found to be in reasonable semiquantitative agreement with the dephasing length $L_\varphi$ extracted from the weak-localization magnetoresistance studies. This work demonstrates that the UCF effect is particularly pronounced in self-assembled conducting nanowires. Studies in this direction using metallic nanowires may thus provide insightful information on the UCF mechanism in miniature conductors.

\begin{acknowledgments}

The authors are grateful to S. P. Chiu and Y. H. Lin for valuable experimental assistance and discussion, C. S. Chu for useful discussion, and F. R. Chen and J. J. Kai for providing us with the ITO nanowires used in this study. This work was supported by Taiwan National Science Council through Grant Nos. NSC 99-2120-M-009-001 and NSC 100-2120-M-009-008, and by the MOE ATU Program.

\end{acknowledgments}

\appendix\section{One-dimensional weak-localization magnetoresistance}

The magnetoresistance in the weak-localization effect has been well established over the years, both theoretically \cite{Altshuler-rev87} and experimentally. \cite{Lin-jpcm02,Bergmann-pr84} Our measured low-field MR's in every NW have been least-squares fitted to the 1D WL theoretical prediction as given below ($B$ is applied perpendicular to the NW axis): \cite{Altshuler-rev87,Hsu-prb10,Niimi-prb10}
\begin{widetext}
\begin{eqnarray}
\frac{\triangle R(B)}{R(0)} & = & \frac{e^2}{\pi \hbar} \frac{R}{L} \Biggl\{
\frac{3}{2} \Biggl[ \Biggl( \frac{1}{L_\varphi^2} + \frac{4}{3L_{so}^2} +
\frac{W^2}{3 L_B^4} \Biggr)^{-1/2} - \Biggl(
\frac{1}{L_\varphi^2} + \frac{4}{3L_{so}^2} \Biggr)^{-1/2} \Biggr] -
\frac{1}{2} \Biggl[ \Biggl( \frac{1}{L_\varphi^2} + \frac{W^2}{3 L_B^4}
\Biggr)^{-1/2} - L_\varphi \Biggr] \Biggr\} \,, \label{A1}
\end{eqnarray}
\end{widetext}
%\begin{eqnarray}
%\frac{\triangle R(B)}{R(0)} & = & \frac{e^2}{\pi \hbar} \frac{R}{L} \Biggl\{
%\frac{3}{2} \Biggl[ \Biggl( \frac{1}{L_\varphi^2} + \frac{4}{3L_{so}^2} +
%\frac{W^2}{3 L_B^4} \Biggr)^{-1/2} \nonumber \\ & - & \Biggl(
%\frac{1}{L_\varphi^2} + \frac{4}{3L_{so}^2} \Biggr)^{-1/2} \Biggr] -
%\frac{1}{2} \Biggl[ \Biggl( \frac{1}{L_\varphi^2} + \frac{W^2}{3 L_B^4}
%\Biggr)^{-1/2} - L_\varphi \Biggr] \Biggr\} \,, \label{A1}
%\end{eqnarray}
where $\triangle R(B) = R(B) - R(0)$, $R$ is the resistance of a nanowire of width $W$ and length $L$, $L_B = \sqrt{\hbar /eB}$ is the magnetic length, $L_\varphi = \sqrt{D \tau_\varphi}$ is the electron dephasing length, $L_{\rm so} = \sqrt{D \tau_{\rm so}}$ is the spin-orbit scattering length ($\tau_{\rm so}$ being the spin-orbit scattering time), and the electron diffusion constant $D = v_F^2 \tau_e/3$ ($v_F$ being the Fermi velocity, and $\tau_e$ being the electron elastic mean free time). Notice that our NW's are 1D with regard to the WL effect (i.e., $d < L_\varphi$), while 3D with regard to the usual Boltzmann transport (i.e., $l \ll d$). The spin-orbit scattering length (time) is a temperature independent quantity whose size, relative to the inelastic electron scattering strength, determines the sign of the weak-(anti)localization effects in the low-field MR. \cite{Altshuler-rev87} Our method for estimating the $v_F$, $\tau_e$, and $D$ values in the ITO NW's through our measured Fermi energy $E_F$ values in the ITO material has recently been described in Ref. \onlinecite{Hsu-prb10}.

\begin{figure}
\includegraphics[width=10cm,height=8cm]{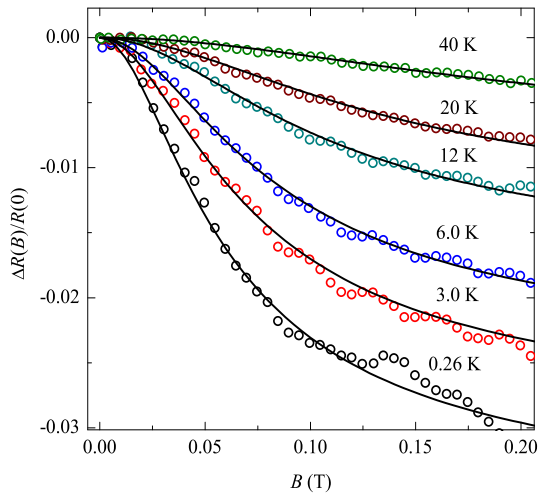}
\caption{\label{1DMR} Normalized magnetoresistance as a function of perpendicular magnetic field of the NW14 nanowire at several temperatures, as indicated. The symbols are the experimental data and the solid curves are the theoretical predictions of Eq.~(\ref{A1}). Note that the theory can well describe the experimental data. Note also that signatures of the UCF's in $B \gtrsim 0.1$ T can be seen especially at 0.26 and 3.0 K.}
\end{figure}

Figure~\ref{1DMR} shows the normalized magnetoresistance $\triangle R(B) /R(0)$ as a function of magnetic field at several temperatures of the NW14 nanowire at first cooldown. The symbols are the experimental data and the solid curves are the theoretical predictions of Eq.~(\ref{A1}). This figure clearly demonstrates that our measured low-field MR's can be well described by the 1D WL theory in the wide temperature interval of 0.26--40 K. Therefore, the characteristic electron dephasing length $L_\varphi$ can be very reliably extracted. The $L_\varphi$ values thus obtained are plotted in Fig.~\ref{fig8}(a). Similarly, our measured low-field MR's in the NW12 nanowire can also be well described by Eq.~(\ref{A1}) (not shown), and the extracted $L_\varphi$ values are plotted in Fig.~\ref{fig8}(b). Besides, our measured low-field MR's of the NW28 nanowire, along with the inferred $L_\varphi$ values, have previously been reported in Ref. \onlinecite{Hsu-prb10}. Finally, our extracted spin-orbit scattering length is $L_{\rm so} \simeq 125$ nm in the NW28 nanowire. On the other hand, the spin-orbit scattering is comparatively weak in the relatively cleaner NW12 and NW14 nanowires, and hence only a lower bound can be estimated, i.e., $L_{\rm so} \gtrsim 0.5$ $\mu$m (corresponding to $\tau_{\rm so} \gtrsim 300$ ps) in these two NW's. We would like to note that a weak spin-orbit scattering strength has recently also been found in a series of homogeneous and inhomogeneous ITO thin films. \cite{Wu-prb}


\begin{thebibliography}{99}
\bibitem{thermal} In our recent studies of the temporal UCF's in RuO$_2$ NW's due to mobile defects (Ref. \onlinecite{Lien-prb11}), we had compared our measured rms UCF magnitudes with the theoretical predictions in the absence of the thermal averaging effect [i.e., a factor of $(L_T/L_\varphi^{\rm UCF})^2$ is not multiplied to the Eqs. (1) and (2) of Ref. \onlinecite{Lien-prb11}]. A good agreement was obtained. In that case, it is also not understood why the thermal averaging effect played no significant role. It should be of fundamental importance to reexamine whether any ingredients in the theoretical concepts for thermal averaging in mesoscopic physics might have been overlooked (overestimated).
\bibitem{uncertainty}  Around 10 K, the experimental magnitudes of $F(0)$ are relatively small, rendering the evaluated $B_c$ and $L_\varphi^{\rm UCF}$ values less accurate compared with those extracted for $T \lesssim 5$ K. Moreover, the uncertainty in $d$ could introduce some error in $L_\varphi^{\rm UCF}$.
\bibitem{e-ph} The electron-phonon relaxation rate, which often dominates in normal metals at a few degrees K and higher, is relatively weak in the low-carrier-concentration ITO material, and thus can be ignored, see Ref. \onlinecite{Wu-prb}.
\end{thebibliography}
\end{document}